\documentclass[notitlepage,aps,prd,groupedaddress,nofootinbib]{revtex4-1}
\usepackage[sc]{mathpazo} % Use the Palatino font
\usepackage[T1]{fontenc} % Use 8-bit encoding that has 256 glyphs
\usepackage{microtype} % Slightly tweak font spacing for aesthetics

\usepackage[hmarginratio=1:1,top=32mm,columnsep=20pt]{geometry} % Document margins
\usepackage{booktabs} % Horizontal rules in tables
\usepackage{float} % Required for tables and figures in the multi-column environment - they need to be placed in specific locations with the [H] (e.g. \begin{table}[H])
\usepackage{hyperref} % For hyperlinks in the PDF
\usepackage{graphics}
\usepackage{graphicx}
\usepackage{float}
\usepackage{amssymb}
\usepackage{slashed}
\usepackage{fancybox}
\usepackage{natbib}
\usepackage{amsfonts}
\usepackage{epsfig}
\usepackage{rotating}
\usepackage{amsmath}
\usepackage{wasysym}%
\usepackage{braket}
\usepackage{epsfig}
\usepackage{lettrine} % The lettrine is the first enlarged letter at the beginning of the text
\usepackage{paralist} % Used for the compactitem environment which makes bullet points with less space between them

\usepackage{titlesec} % Allows customization of titles
\renewcommand{\thesection}{\Roman{section}}
\renewcommand{\thesubsection}{\thesection.\Roman{subsection}}
\titleformat{\section}[block]{\large\scshape\centering}{\thesection.}{1em}{} % Change the look of the section titles
\titleformat{\subsection}[block]{\large}{\thesubsection.}{1em}{} % Change the look of the section titles

\usepackage{fancyhdr} % Headers and footers
\allowdisplaybreaks

\begin{document}

%\preprint{}

\title{Grand unified hidden-sector dark matter}


\author{Stephen J. Lonsdale, Raymond R. Volkas}
%\thanks{}
\affiliation{ARC Centre of Excellence for Particle Physics at the Terascale, School of Physics,\\ The University of Melbourne, Victoria 3010, Australia}

%\date{\today}
\begin{abstract}
We explore $G\times G$ unified theories with the visible and the hidden or dark sectors paired under a $\Bbb{Z}_2$ symmetry. 
Developing a system of 'asymmetric symmetry breaking' we motivate such models on the basis of their ability to generate 
dark baryons that are confined with a mass scale just above that of the proton, as motivated by asymmetric dark matter. 
This difference is achieved from the distinct but related confinement 
scales that develop in unified theories that have the two factors of $G$ spontaneously breaking in an asymmetric manner.  We show how Higgs potentials that admit different 
gauge group breaking chains in each sector can be constructed, and demonstrate the capacity for generating different fermion mass scales. Lastly we discuss supersymmetric extensions of such schemes.
\end{abstract}

%\keywords{}

\maketitle

%\section{}
%\subsection{}
%\subsubsection{}

\section{ \bf  Introduction }\label{intro}
%\noindent{
Observations have established that our universe is composed of $\sim 32 \%$ matter and 
$\sim 68\% $ dark energy. Of the matter, only $\sim 15 \% $ is accounted for by the particles 
that make up the standard model. 
The make-up of the remaining dark matter (DM) is one of the chief 
concerns of present day physics. The visible matter (VM) is 
composed of three generations of quarks and leptons interacting under SU(3) $\times$ SU(2) $\times$ U(1) 
gauge interactions, plus a Higgs boson. 
It is common to consider that DM may be a similar set of 
particles charged under a different gauge group with only limited interactions 
with ordinary matter. The two sectors, the visible 
and the dark, provide an explanation for why evidence of DM has only been 
encountered so far through gravitational effects and the question of how these sectors
could form to be so separate is an interesting challenge.

Asymmetric dark matter models, a broad category within hidden-sector scenarios, relate the creation of the mass density in the visible 
sector to the generation of matter in the dark sector. 
The fact that the mass densities of DM and VM in the universe are 
seen to be of the same order,
\begin{equation}
 \Omega_{DM} \simeq 5\Omega_{VM},
\end{equation}
suggests that the mechanism by which VM was created in the early universe
is connected to the production of DM. 
The established origin of the relic density of VM relies on a baryon asymmetry, in 
which a small excess of baryons over antibaryons
developed, and after the antibaryons had all annihilated with opposing baryons
only a baryon density remained. 
In asymmetric dark matter models the asymmetry in each sector is connected by the 
conservation of a global quantum number. Once the symmetric parts 
in each sector have annihilated away then the number densities of the remaining 
particles in each sector are related to each other \cite{Davoudiasl:2012uw, Petraki:2013wwa, Zurek:2013wia}.
However Eq.1 is a mass-density relation. In order to explain it, a theory of how the DM mass is related 
to the proton mass is needed in addition to related number densities.
Now, grand unified theories (GUTs) unite the fundamental forces of particle physics into a single gauge group at high energy along with their 
coupling constants. The purpose of this paper is to explore how GUTs can relate a dark-sector confinement scale to the QCD scale.

We approach this by demonstrating an 'asymmetric symmetry breaking' mechanism in
which isomorphic and $\Bbb{Z}_2$ related gauge groups $G_V \times G_D$ of the visible and dark sectors naturally differ from each other after symmetry breaking. 
Each sector then features different mass scales for visible and dark baryons. 
We now briefly review how this mass is generated in our own sector.

%}

%	PROTON/Baryon mass
%----------------------------------------------------------------------------------------

%\noindent{ 
The dependence of the running coupling constant of QCD, $  \alpha_s(\mu)  $, on the scale $ \mu$ can be 
expressed in two ways. 
The first is as a function of a reference scale $\mu_0$ which gives an equation of the 
form
\begin{equation}
 \alpha_s(\mu) = \frac{\alpha_s(\mu_0)}{1+ (\beta_0/4\pi)\alpha_s(\mu_0)\ln(\mu^2/{\mu_0}^2)},
\end{equation}
 where $\alpha_s$ is known at the reference scale. 
Alternatively the dependence can be expressed as 
\begin{equation}
  \alpha_s(\mu) = \frac{4\pi}{\beta_0 \ln(\mu^2/ \Lambda^2)},
\end{equation}
in which the parameter $\Lambda$ is the confinement scale, the value at which 
the strong coupling constant becomes large as the energy scale decreases. 
This is a distinct feature of asymptotic freedom in which $\beta_0 > 0$ . At first order the beta function for SU(3) is 
\begin{equation}
 \beta_0 = 11-\frac{2}{3}n_f ,
\end{equation}
where $n_f$ is the number of quark flavours that appear in the loop corrections 
at a given energy scale.
If one then knows the value of the strong coupling constant at a high energy scale
$U$, for instance at a GUT scale, it is possible to calculate the value
of the confinement 
scale by evolving the coupling constant and taking into account quark mass thresholds.
The threshold values are actually at twice the mass of each quark as this is 
the amount of energy needed to switch on the relevant loop correction.
The resulting equation is dependent on this high reference scale, $U$,  $\alpha_s$
at said scale, 
and the masses of the fermions in the range between the two scales. One obtains
\begin{equation}
 \Lambda = 2^{2/9}  e^{-2\pi /9 \alpha_s(U)}  U^{\frac{7}{9}} m_c^{\frac{2}{27}} m_b^{\frac{2}{27}} m_t^{\frac{2}{27}} .
\end{equation}
where $m_{t,b,c}$ are the top-, bottom-, and charm-quark masses. For a more general theory the confinement scale  is given by
\begin{equation}
 \Lambda = 2^{  1-\frac{b_u}{b_c} }  e^{-2\pi / \alpha_s(U)b_c}  U^{\frac{b_u}{b_c}} m_1^{\frac{b_c-b_b}{b_c}} m_3^{\frac{b_t-b_u}{b_c}}m_2^{\frac{b_b-b_t}{b_c}}.
\end{equation}
The terms labeled $b_x$ in this form of the equation denote the values of $\beta_0$ for 
different numbers of contributing  quark flavours. For instance, $b_b$ is the value above 
twice the charm mass but below the bottom mass. We use this notation for the 
sake of the more generalised relationship between energy thresholds and the DM confinement scale where the number of 
massive quarks and the masses that they have are initially completely free parameters. 
Only the masses of quarks larger than $\Lambda$ itself appear explicitly in the equation. 
It is important to note that this equation is very sensitive to the value of
the scale $U$. This sensitivity is avoided, however, in a non-abelian dark sector if the confining gauge group is also SU(3), as we explain below.
To form the alternate gauge groups we develop a systematic way of generating 
different dark sectors from unified origins, with both containing an unbroken SU(3) factor.

The idea of connecting DM to unified origins is not new, of course, and a  large number of 
models explore the possibility of DM coming from a dark sector which closely resembles 
our own. In particular our work is related to the theory 
of mirror matter \cite{Lee:1956qn,Kobzarev:1966,Pavsic:1974rq,Blinnikov:1982eh,Blinnikov:1983,Foot:1991bp,Foot:1991py,Foot:1995pa,Berezhiani:1995yi,Foot:2000tp,Berezhiani:2000gw,Ignatiev:2003js,Foot:2003jt,Foot:2004pq,Berezhiani:2003wj,Ciarcelluti:2004ik,Ciarcelluti:2004ip,Foot:2014mia}, 
where the two sectors share 
the same gauge group and the hidden sector is an exact copy of the standard model. Our work demonstrates
the capacity for natural symmetry breaking of mirrored GUT groups to two sectors which are manifestly different at 
low energies in both gauge symmetry and the masses of their particle content. 
Where a number of other works, in particular \cite{Bai:2013xga,Barr:2013tea,Ma:2013nga,Newstead:2014jva}, posit the existence of a hidden non-abelian 
gauge group 
responsible for generating dark baryon mass, or in the case of \cite{Boddy:2014yra}, glueball mass, we aim to show that such 
confining dark sector groups 
and in particular SU(3) can appear spontaneously from a unified original gauge symmetry and that the 
generation of a confinement scale in the dark sector which is different from (often larger than) our own can be a natural consequence of
the way in which these unified theories break 'asymmetrically'.

Recently \cite{Tavartkiladze:2014lla} explored models of composite fermions from SU(5)$\times$SU(5) with a discrete symmetry that used 
potentials with different symmetry breaking scales to achieve coupling unification and generate confining particles 
in the TeV range. The intention of this work is to explore the broader possibilities of generating spontaneous differences in $G_V \times G_D$ 
theories to answer why DM could have a mass of the same order as VM. 

The next section gives the results on the dark SU(3) confinement scale as a function of dark-quark mass in non-supersymmetric theories.  Section \ref{sec:ASB} then explains the basic idea behind asymmetric symmetry breaking in the simplest possible context.  With that as a springboard, Sec.~\ref{sec:SU5xSU5} shows how the mechanism can be implemented in non-supersymmetric SU(5), while Sec.~\ref{sec:FM} deals with how dark-quark mass generation can be automatically different from ordinary quark mass generation.  Section \ref{sec:SO10xSO10} briefly discusses the rather different and diverse possibilities afforded in SO(10) constructions, and Sec.~\ref{sec:pheno} touches on phenomenological constraints.  Attention then turns to supersymmetry, with Sec.~\ref{sec:SUSYASB} showing how asymmetric symmetry breaking can be implemented, and Sec.~\ref{sec:SUSYDarkSU3} displaying the dark confinement scale results.  Final remarks are made in Sec.~\ref{sec:conc}.  Appendices A and B give further details of the scalar potential analyses in the non-SUSY and SUSY case, respectively.

\section{\bf Dark SU(3)}\label{sec:darkSU3}
The goal of these models is to obtain the standard model 
in one sector and a naturally occurring but distinctly different dark sector 
with its own SU(3) gauge group that facilitates an asymptotically free strong binding of 
dark quarks into heavy dark baryons. These baryons could then account for the relative mass density difference between the total 
visible and DM in a universe where the number density generation is governed
by asymmetric dark matter dynamics.

In a model with unified coupling constants, and where at a high energy the gauge groups of 
each sector break to SU(3) at the same scale, the two values of the 
strong coupling constant $\alpha_s $ and $\alpha_{s_D} $ are the same 
from the GUT breaking scale all the way down to the scale at which the number of 
possible fermions in the loop corrections first deviates between the two sectors or further
symmetry breaking occurs.
This is highly desirable as it allows the equation of the dark confinement scale to be
greatly simplified, as the high reference scale can then be chosen to be at the 
value of  $\alpha_{s_D}$,  the deviation point which has just the value of the standard model 
$\alpha_s$ at the scale of either the top quark or the heaviest of the dark quarks 
depending on which of these two has greater mass.
If we make the further 
assumption for the sake of simplicity
that all heavy dark quarks have the same mass, then our equation becomes a function 
of just one continuous and one discrete parameter, namely the dark fermion mass scale $m$
and the number of fermions, $n_f$, that are at 
such a scale, $\Lambda(n_f, m)$. 
\begin{figure}[t!]
\centering
\includegraphics[angle=0,width=0.7\textwidth]{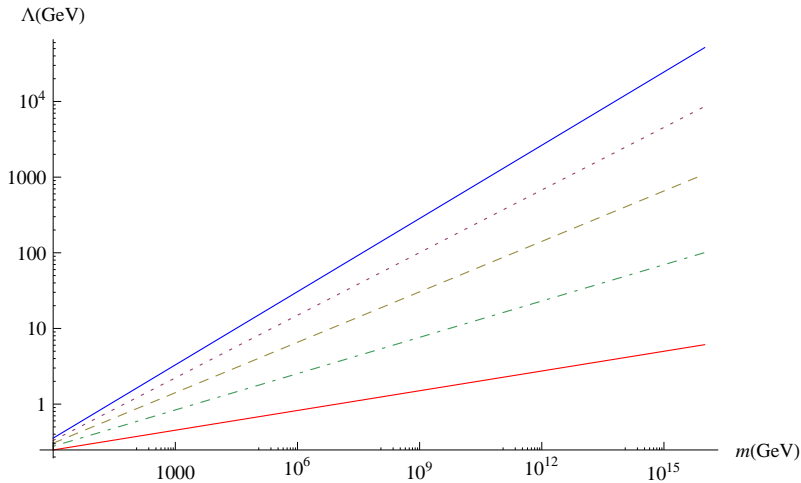}
\caption{Confinement scale dependence on fermion masses with five heavy and one
light quark at the top through five light and one heavy quark at the bottom line. 
All heavy quarks have mass $m$ and all light quarks are taken to have masses below $\Lambda$.}
\end{figure}
Figure 1 shows the relation between the dark confinement scale and the mass of the heavy quarks
which couple to the dark strong force. One could devise scenarios in which some of the heavy fermions attain an intermediate  mass scale and adjust
the confinement scale accordingly. 
Figure 1 shows that a value of $\Lambda$ at approximately one order of magnitude greater 
than the standard model is compatible with dark quark masses up to 1000 TeV. 
The baryons themselves form from the light, or massless, quarks 
and therefore have mass either almost or totally dominated by the confinement scale.
If one can build a model of two such sectors that allows for the dark sector to
give masses to coloured fermions at a low enough energy scale then accordingly one can provide an explanation for the similarity 
in the mass densities of visible and dark matter.

The focus of this paper is on how asymmetric symmetry breaking patterns from $G \times G$ GUTs
can be induced and 
how quark and dark-quark mass generation may naturally differ, as these are the most important
ingredients for determining the dark QCD confinement scale.  Other important features for asymmetric DM models, such as the
asymmetry generation and transfer mechanism and the annihilation of the symmetric part, plus various issues associated with constructing 
fully realistic GUTs,\footnote{In particular, the problem of unsuccessful quark-lepton mass relations requires a  non-minimal scalar sector.  This will not affect our results provided that all coloured components of these multiplets receive GUT-scale masses, as is required in any case to solve the doublet-triplet splitting problem.} are left for future work.

\section{ \bf  Asymmetric Symmetry Breaking}\label{sec:ASB}
In order to illustrate the range of possible asymmetric symmetry breaking models
and explain the basic features that drive asymmetric symmetry breaking
we examine in this section a simple toy model that involves all of the most 
basic terms required and demonstrate 
what vacuum expectation value (VEV) patterns are possible. 
%We also examine the range of possible masses in 
%systems which incorporate multiple energy scales derived from this symmetry breaking.

The simple model we use for illustration is based on four real scalars in two $\Bbb{Z}_2$ pairs,
\begin{equation}
\phi_1 \leftrightarrow \phi_2 ,   \qquad   \chi_1   \leftrightarrow \chi_2.
\end{equation}
The general potential can be written without loss of generality as  
\begin{equation} 
\begin{split}  
V= \lambda_{\phi}(\phi_1^2 + \phi_2^2 - v_{\phi}^2)^2 + \\
 \lambda_{\chi}(\chi_1^2 + \chi_2^2 - v_{\chi}^2)^2 + \\
 \kappa_{\phi} (\phi_1^2 \phi_2^2) + \\
 \kappa_{\chi} (\chi_1^2 \chi_2^2) + \\ 
 \sigma(\phi_1^2\chi_1^2 + \phi_2^2\chi_2^2)+ \\
 \rho(\phi_1^2 + \chi_1^2 + \phi_2^2 + \chi_2^2 - v_{\phi}^2- v_{\chi}^2 )^2.
\end{split}
\end{equation}
Terms such as $\phi_1^3 \phi_2 + \phi_1\phi_2^3$ etc.\ are taken to be absent because of additional discrete symmetries.
If each of the parameters is positive, then each of the six terms in this potential is positive definite. 
Then each is individually minimised if it is equal to zero. 
The first four terms are thus minimised by the condition that for each $\Bbb{Z}_2$ pair, one field gains a nonzero VEV while its partner has strictly zero VEV.
The fifth term is minimised by the condition that the two nonzero-valued fields do not share a subscript (sector). 
The last term is then already zero by the previous conditions and the entire potential is minimised by these
'asymmetric' configurations: 
\begin{eqnarray}
 \braket{\phi_1} =  v_\phi ,   \qquad  \braket{\chi_1} =  0, \nonumber\\           
 \braket{\phi_2} =  0,    \qquad \braket{\chi_2} =  v_\chi .
\end{eqnarray}
Note that it could have been ($\phi_2,\chi_1$) that gained nonzero VEVs, i.e.\ we cannot know \emph{a 
priori} which way the symmetry will break.

A key feature of these asymmetric models is the ability of one asymmetry to induce further
asymmetry in additional $\Bbb{Z}_2$-related fields. If we take a second set of four fields just as in the above case,
\begin{equation}
\Omega_1 \leftrightarrow \Omega_2 ,    \qquad    \eta_1   \leftrightarrow \eta_2,
\end{equation}
our new general potential can be written in the form,
\begin{align} 
%\begin{split}  
V &= \lambda_{\phi}(\phi_1^2 + \phi_2^2 - v_{\phi}^2)^2 + 
 \lambda_{\chi}(\chi_1^2 + \chi_2^2 - v_{\chi}^2)^2 + \kappa_{\phi} (\phi_1^2 \phi_2^2) + 
 \kappa_{\chi} (\chi_1^2 \chi_2^2) \nonumber\\
 &+ \sigma(\phi_1^2\chi_1^2 + \phi_2^2\chi_2^2)+ 
 \rho(\phi_1^2 + \chi_1^2 + \phi_2^2 + \chi_2^2 - v_{\phi}^2- v_{\chi}^2 )^2 \nonumber\\
 \nonumber\\
 &+ \lambda_{\Omega}(\Omega_1^2 + \Omega_2^2 - v_{\Omega}^2)^2 + 
 \lambda_{\eta}(\eta_1^2 + \eta_2^2 - v_{\eta}^2)^2 + \kappa_{\Omega} (\Omega_1^2 \Omega_2^2) + 
 \kappa_{\eta} (\eta_1^2 \eta_2^2) \nonumber\\ 
 &+ \sigma_1(\Omega_1^2\eta_1^2 + \Omega_2^2\eta_2^2)+ 
 \rho_1(\Omega_1^2 + \eta_1^2 + \Omega_2^2 + \eta_2^2 - v_{\Omega}^2- v_{\eta}^2 )^2 \nonumber\\
 \nonumber\\
 &+ \sigma_2(\Omega_1^2\chi_1^2 + \Omega_2^2\chi_2^2) +       
 \rho_2(\Omega_1^2 + \chi_2^2 + \Omega_2^2 + \chi_1^2 - v_{\Omega}^2 - v_{\chi}^2)^2 \nonumber\\  %has a non zero
 \nonumber\\
&+ \sigma_3(\phi_1^2\eta_1^2 + \phi_2^2\eta_2^2) +            
\rho_3(\phi_1^2 + \eta_2^2 + \phi_2^2 + \eta_1^2  - v_{\eta}^2  -v_{\phi}^2 )^2  \nonumber\\ %has a non zero
\nonumber\\
&+ \rho_4(\Omega_1^2 + \phi_1^2 + \Omega_2^2 + \phi_2^2 - v_{\Omega}^2  -v_{\phi}^2 )^2 +       
\sigma_4(\Omega_1^2\phi_2^2 + \Omega_2^2\phi_1^2) \nonumber\\  %0
\nonumber\\
&+ \rho_5(\chi_2^2 + \eta_2^2 + \chi_1^2 + \eta_1^2  -  v_{\eta}^2  -v_{\chi}^2)^2   +      
\sigma_5(\chi_2^2  \eta_1^2 + \chi_1^2  \eta_2^2). %has a non zero \\
%\end{split}
\end{align}
As before, with each term positive definite, the potential is minimised for the following pattern of VEVs:
\begin{eqnarray}
\braket{\phi_1} =  v_\phi\ ,    \qquad  \braket{\chi_1} =  0\ , \nonumber\\
\braket{\phi_2} =  0\ ,    \qquad \braket{\chi_2} =  v_\chi\ , \nonumber\\
\braket{\Omega_1} =  v_\Omega\ ,   \qquad  \braket{\eta_1} =  0\ , \nonumber\\            
\braket{\Omega_2} =  0\ ,   \qquad \braket{\eta_2} =  v_\eta\ . 
\end{eqnarray}
As usual this vacuum is degenerate with its  $\Bbb{Z}_2$ transform.
The potential has been constructed in such a way that the minima are when nonzero $\phi, \Omega$ VEVs share a sector, and the same is true for $\chi$, $\eta$.
This associated asymmetry allows us to link together particular
subgroups from gauge symmetry breaking with appropriate
Higgs multiplets for that specific sector to give different masses to fermions. This idea will be explored further in Sec.~V. 
Large systems of many representations of scalar fields can take an initially mirrored GUT group and naturally
populate each 
sector with nonzero VEVs of different scales which are given to different representations thus making the 
two sectors highly 
divergent in their features though identical in their origins.  This toy model will serve as a proof of concept for the more involved scenarios that we move on to, that is, replacing 
these singlet fields with representations of GUT groups.

% SU(5) specific models
%-------------------------------------------------0---------------------------------------------

\section{\bf SU(5) $\times$ SU(5) Asymmetric Symmetry Breaking}
\label{sec:SU5xSU5}
We now consider how an asymmetric VEV structure allows for
separate mechanisms to generate fermion masses in each sector. 
This section explores an illustrative model of asymmetrical symmetry
breaking that uses the SU(5) GUT candidate.
Paired with a discrete symmetry our $SU(5)_v \times SU(5)_d$ will be broken to different 
gauge groups in the two sectors but with both featuring unbroken SU(3) subgroups which have quantitative 
differences.
This then allows a numerical difference in the value of the dark sector confinement scale. To 
accomplish this we build a symmetry breaking potential out of four scalar multiplets making 
use of two different representations of SU(5), 
namely the  \textbf{24}  and the  \textbf{10}, each of which will have one of two multiplets become the sole 
attainer of a nonzero VEV in just one sector thus facilitating the different symmetry
breaking patterns.
In its most basic form this is just an extension of the simple model of Sec.~III in which 
the two sectors are the visible and dark and the fields $\phi_1$, $\phi_2$ are now \textbf{24}
dimensional multiplets while $\chi_1$, $\chi_2$ become two copies of the \textbf{10} representation of SU(5),
\begin{eqnarray}
\phi_v \sim (24,1)\ ,  \qquad \chi_v \sim (10,1)\ ,\nonumber\\
\phi_d \sim (1,24)\ ,  \qquad \chi_d \sim (1,10)\ .
\end{eqnarray}
Consider firstly the \textbf{10} representation of SU(5) which one uses to spontaneously break 
\begin{equation}
 SU(5)_d \rightarrow SU(3)\; \times \;SU(2)
\end{equation}
by appropriate choice of the sign of parameters in a general quartic
scalar potential. The general renormalisable potential for a scalar multiplet $\chi \sim 10$ is ,
\begin{equation}
 V_{10} = -\mu_t^2 \chi_{ij}\chi^{ji} + \lambda_{t1} (\chi_{ij}\chi^{ji})^2 + \lambda_{t2} \chi_{ij}\chi^{jk}\chi_{kl}\chi^{li}.
\end{equation}
Note that $i,j={1,...,5}$ are SU(5) gauge indices with $\chi_{ij}=-\chi_{ji}$, and the subscript $t$ denotes `ten'.
Choosing the parameter $\lambda_{t2}$ to be 
negative produces a VEV that breaks SU(5) to SU(3) $\times$ SU(2) \cite{Li:1973mq}.

In the other sector the method of breaking SU(5) to the standard model is to use scalar
fields in the adjoint representation. The quartic potential is
\begin{equation}
 V_{24} = -\mu_a^2 \phi^i_j\phi^j_i + \lambda_{a1} (\phi^i_j\phi^j_i)^2 + \lambda_{a2} \phi^i_j\phi^j_k\phi^k_h\phi^h_i,
\end{equation}
where the subscript $a$ is for `adjoint', and $\phi$ is Hermitian traceless. Choosing $\lambda_{a2}$ to be positive gives us the breaking 
\begin{equation}
 SU(5)_v \rightarrow SU(3) \; \times \; SU(2) \; \times \; U(1).
\end{equation}
In this model we have four representations of scalar fields in the two $\Bbb{Z}_2$ pairs of Eqs.\ 15 and 16.
The complete, general fourth-order, gauge-invariant scalar potential invariant under the discrete 
symmetry is written in Appendix A. It contains two copies of each of the above 
two potentials for the multiplets in each sector as well as all possible gauge-invariant contractions between 
the  \textbf{24}  and \textbf{10} in each individual sector, that is, 
of the style ${\chi_v} {\chi_v} {\phi_v} {\phi_v}$.

We can take these basic potentials written above and use them to write a simple outline of 
the full potential. 
We first duplicate each of the above potentials to accommodate each one's dark counterpart, 
and add in the cross terms such as $Tr(\phi_v^2)Tr(\phi_d^2)$. We term these

\begin{equation}
V_A = V_{24} + V_{24}' + \kappa_a Tr[{\phi_v^2}] Tr[{\phi_d^2}] 
\end{equation}
and 
\begin{equation}
 V_ T= V_{10} + V_{10}' + \kappa_t {\chi_v}_{ij} {\chi_v}^{ji}{\chi_d}_{nm} {\chi_d}^{mn}.
\end{equation}
To this there are five remaining contractions that we must add to write 
the general renormalisable potential. A portion of this potential, displayed in full in Appendix A, can then be written as
\begin{equation}
 V= V_A + V_T  +
C_0( {\chi_d}_{nm} {\chi_d}^{mn} Tr[\phi_v^2] + 
{\chi_v}_{ij} {\chi_v}^{ji} Tr[\phi_d^2]) 
+ \ldots
\end{equation}
Extending the analysis of Sec.~III we find that for a 
particular region of parameter space in this potential, the global 
minimum is at
\begin{eqnarray}
\braket{\phi_v} & =  & v_v
\begin{pmatrix}
  1 & 0 & 0 & 0 & 0 \\
   0 & 1 & 0 & 0 & 0 \\
 0 & 0 & 1 & 0 & 0 \\
  0 & 0 & 0 & -3/2 & 0 \\
   0 & 0 & 0 & 0 & -3/2 \\
 \end{pmatrix}, \nonumber\\ 
\braket{\chi_v} & = & 0, \nonumber\\
\braket{\phi_d} & = & 0, \nonumber\\
\braket{\chi_d} & = & v_d
 \begin{pmatrix}
  0 & 1 & 0 & 0 & 0 \\
  -1 & 0 & 0 & 0 & 0 \\
  0 & 0 & 0 & 0 & 0 \\
  0 & 0 & 0 & 0 & 0 \\
  0 & 0 & 0 & 0 & 0 \\
 \end{pmatrix}.
\end{eqnarray}
By using the principles of the simple model and its parameter space from Sec.~III, this 
potential is seen to induce the two SU(5) gauge groups to indeed break differently in 
each sector. 
In one sector the \textbf{10} representation attains a VEV breaking SU(5) to SU(3) $\times$ SU(2) and the 
positive definite contraction terms push the \textbf{24} in that sector to attain a VEV of zero. 
In the other sector the \textbf{10} representation is driven to have a VEV of zero by contraction 
terms with its counterpart and this forces the \textbf{24} to attain a VEV that breaks this second 
SU(5) to the standard model gauge group. 
There is once again no way of knowing which is the visible and which is the dark sector 
prior to symmetry breaking. 
Once the symmetry is broken to the lowest state it shall simply be that we label the SU(5) 
which is broken to the standard model group the gauge 
symmetry of the visible sector and the alternatively broken symmetry is then the dark
sector gauge group. 

We now explore fermion mass generation with a view to having the visible and dark colour SU(3) gauge coupling constants evolve differently under the renormalisation group.

%Fermions masses in an SU(5)
%------------------------------------------------

\section{\bf Fermion Masses}
\label{sec:FM}
In SU(5) theories the fermions of the standard model are assigned to
the $\mathbf{\overline{5}}$ and \textbf{10} representations. 
The product of these allows for mass generation through Yukawa couplings to 
Higgs fields in \textbf{5},  \textbf{10}, \textbf{45} or \textbf{50} dimensional representations. As an example, we aim to have 
two different representations for our mass generation, a \textbf{5} to accommodate the 
standard model Higgs doublet in the visible sector
and another representation which attains a nonzero VEV in the dark sector to give 
a different form of mass generation for the dark sector quarks.\footnote{The idea of a non-abelian gauge sector responsible for confining DM 
has been detailed in a number of different works such as \cite{Bai:2013xga} in which the 
range of SU(N) groups and ultraviolet boundary conditions of the coupling constants that allow for 
TeV-scale-confined DM were investigated. In \cite{Boddy:2014yra} the scale 
of gluinos and glueballs in an SU(N) hidden sector was seen to be adjustable to 
produce TeV scale glueball DM that could 
agree with a number of astrophysical constraints of self-interacting DM. }

The \textbf{10} representation
already employed in the 
symmetry breaking 
only gives mass to leptons and is thus unsuitable. We therefore choose to 
examine how a \textbf{5} and a \textbf{45} in each sector can allow for a difference in the scale of quark and dark-quark masses. 
The \textbf{45} has the interesting property of automatically leaving one dark quark massless \cite{Ross:1985ai}, which is a very useful feature for our application.
The fermion multiplets are the same in each sector, again respecting our initial mirror symmetry:
\begin{eqnarray}
\psi_{v_5}\sim (\overline{5},1)\ ,  \qquad \psi_{d_5}\sim (1,\overline{5})\ ,\nonumber\\
\psi_{v_{10}} \sim (10,1)\ ,  \qquad \psi_{d_{10}} \sim (1,10)\ ,
\end{eqnarray}
and the Higgs multiplets which take the place of the fields $\Omega$, $\eta$ from Sec.~III are
\begin{eqnarray}
H_{v_5} \sim (5,1)\ ,  \qquad H_{d_5} \sim (1,5)\ ,\nonumber\\
H_{v_{45}} \sim (45,1)\ ,  \qquad H_{d_{45}} \sim (1,45)\ .
\end{eqnarray}
The Yukawa Lagrangian is
\begin{equation}
 \mathcal{L}_F= y_{1}\overline{\psi_{v_5}} H^{*}_{v_5} \psi_{v_{10}} + y_{2}\overline{\psi_{v_{10}}} H_{v_{45}} \psi_{v_{10}}  + 
 y_{1}\overline{\psi_{d_5}} H^{*}_{d_5} \psi_{d_{10}} + y_{2}\overline{\psi_{d_{10}}} H_{d_{45}} \psi_{d_{10}} + H.c.
\end{equation}
The methodology of Sec.~IV can be extended to include the $\Bbb{Z}_2$ scalar pairs responsible for fermion mass generation.
The asymmetric symmetry breaking described in Sec.~III can induce consecutive asymmetries 
in more sets of fields. 
The dependence for which way the asymmetry in the second set will fall is entirely dependent on the weighting of the cross terms between the two sets.  

It is in this manner that we arrange for the $H_{45}$ in the 
visible sector to have a zero VEV, while in 
the dark sector it gives mass to five of the six quarks at an indeterminate scale 
$v_d$ and reduces the dark sector symmetry from SU(3) $\times$ SU(2) to SU(3). 
The invariant component of $H_{d_{45}}$  is
\begin{equation}
\braket{H_{d_{45}} }^{b5}_{a}= v_d
\begin{pmatrix}
  1 & 0 & 0 & 0 & 0 \\
   0 & 1 & 0 & 0 & 0 \\
 0 & 0 & 1 & 0 & 0 \\
  0 & 0 & 0 & -3 & 0 \\
   0 & 0 & 0 & 0 & 0 \\
 \end{pmatrix}.
\end{equation}
On the other hand the $H_5
$ has a VEV of zero in the dark sector and a nonzero VEV in the visible
sector as per the minimal SU(5) model of giving mass to the fermions:   
\begin{equation}
 \braket{H_v} = v_v (0,0,0,0,1),   \qquad \braket{H_d} = 0.
\end{equation}
The scale $v_d$ can then be compared to the top line in Fig.\ 1 from Sec.~II 
in which we have five heavy dark quarks and a single massless dark quark. 
In such a scenario, if the masses of the quarks are less than 1000 TeV then they produce dark confinement 
scales less than 14 GeV. 
The remaining massless quark, a dark up-quark, forms a set of neutral $\Delta$(uuu) baryon-like states, lighter than all other possible dark colour singlets 
and with mass completely dominated by the confinement scale.  
This forms a dark analogue of the visible sector nucleon but with mass that is an order 
of magnitude greater. 
If we consider minimal differences 
in the magnitude of the mass generating VEVs, which is quite natural to obtain if parameters are of similar order, 
then at around the electroweak scale, $\sim $246 GeV,
 a confinement scale of $ 2.1$ GeV is generated in the dark sector. This is around an order of
magnitude higher than the standard model QCD scale of 0.217 GeV.

\section{ \bf SO(10) $\times$ SO(10)}\label{sec:SO10xSO10}
We now briefly touch on the subject of SO(10) $\times$ SO(10) and other GUT models and their scope 
with regard to asymmetric symmetry breaking. 
In extending the grand unification from SU(5) to SO(10) we open up a number of possible 
pathways to break down to the standard model.
In particular we could consider breaking to the familiar SU(5) $\times$ SU(5) that we showed previously 
or instead use asymmetric symmetry breaking
to take other paths in both sectors towards a final standard model gauge group and non-abelian
dark sector group. For example the use of the \textbf{45} and \textbf{54} representations allows one sector to take the 
Pati-Salam symmetry breaking path \cite{Pati:1973rp}, and the other that of Georgi-Glashow SU(5) \cite{Georgi:1974sy}.
Alternatively one could bypass SU(5) in one or both sectors altogether. The large number of possibilities raises the
prospect of many different ways to adjust the confinement scale for dark QCD. 
In addition, we could consider the possibilities of how we can adjust the scale of breaking to subgroups in each sector.
As a simple example of what we mean we briefly examine a theory which at a high GUT scale $M_G$ breaks as per
\begin{equation}
 SU(5)_v  \rightarrow SU(3) \times SU(2) \times U(1) ,
\label{eq:vtoSM}
\end{equation}
and
\begin{equation}
 SU(5)_d \rightarrow  SU(4)_d ,
\end{equation}
then at an intermediate scale $M_i$ features
\begin{equation}
SU(4)_d  \rightarrow SU(3)_d.
\label{eq:4dto3d}
\end{equation}
Then we can consider how the confinement scale of $SU(3)_d$ changes with the value of $M_i$.
If, by way of a simple example, one assumes that the visible and dark quarks of each sector have the same masses, then the higher values 
of $M_i$ will yield lower confinement scales as the theory will run as SU(3) for a greater energy span.

\begin{figure}[t!]
\centering
\includegraphics[angle=0,
width=0.7\textwidth]{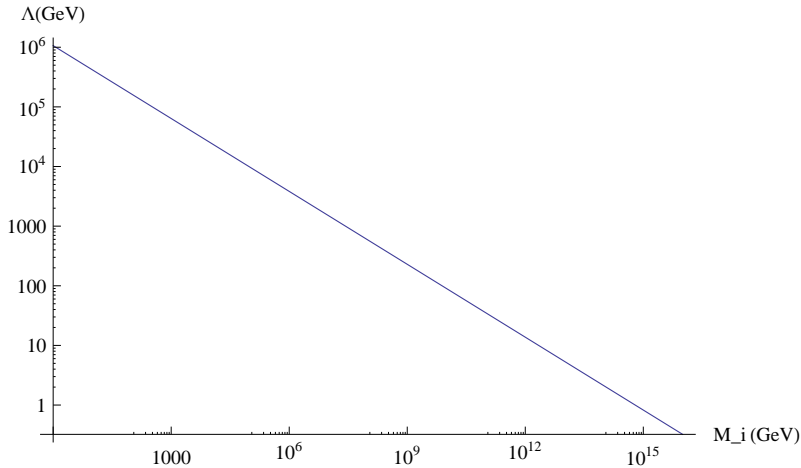}
\caption{The value of the confinement scale for different intermediate breaking scales $M_i$ for the SO(10) $\times$ SO(10) scenario of Eqs.\ \ref{eq:vtoSM}-\ref{eq:4dto3d}. }
\end{figure}

Figure 2 shows that higher values of $M_i$ take us closer to the previous analysis of the standard model. 
This is what one would expect, as in the limit $M_i \rightarrow M_G $ it is as though we have broken 
directly to 
SU(3). If we, however, allow for 
more of the running to be governed by SU(4) then the theory will blow up at a higher scale once we transition to SU(3). 
In SO(10)$\times$ SO(10) models, the many possibilities warrant a dedicated analysis in future work. 

\section{\bf Phenomenological Issues}
\label{sec:pheno}
It is important to note that we merely assumed in the previous analysis that the  
gauge coupling constants of the sectors unify at a high GUT scale.
While the scenario of a non-supersymmetric asymmetric model that we have described
does not automatically have gauge coupling unification, it is possible to bring 
the three coupling constants of the standard model together at the GUT scale by the addition 
of extra Higgs doublets. One must also consider the constraints from the experimental lower bounds of proton 
decay. Decay modes from minimal SU(5) models have quite high 
bounds,  $\tau (p \rightarrow \pi^0 e^{+}) \gtrsim 10^{34}$ years \cite{Senjanovic:2009} and the order
of magnitude estimation for the width 
\begin{equation}
  \Gamma \approx \alpha^2 \frac{m_{p}^{5}}{M_{X}^4},
\end{equation}
demands that we must have at least 
$M_X$ $\approx$ $4 \times 10^{15}$. In \cite{Dorsner:2005ii} it was shown that consistent proton decay limits and unification could 
be obtained with the addition of Higgs multiplets in a non-supersymmetric SU(5). 

Bounds on the dark baryons as DM from the bullet cluster observation are similar to that in \cite{An:2009vq} where the
self-interaction cross section of these
nucleons $\sigma \sim 10^{-26}\;  {\rm cm}^2$ is compared to the upper bound of the DM
self-interaction cross section $\le 10^{-23} \; {\rm cm}^2$ \cite{An:2009vq,Markevitch:2003at,Spergel:1999mh}.
The scale that $v_d$ can take is something that we have not followed in full detail
opting to simply take as a guide the range of scale differences that we can 
accommodate in the simple model in Sec.~III. 
These lead us to see that the scale  of $v_d$ for a factor of five difference 
between ordinary and dark baryons would need to be between $\sim 30$ GeV to  $10^4$ TeV 
depending on how many of the heavy quarks are given mass. The \textbf{45} representation of SU(5) 
would observe the lower bound of $\sim 30$ GeV as the mass scale would give this exact ratio.
If, on the other hand, one only gave mass to a single quark in the dark sector then a 
very high mass would be compatible with a confinement scale of order the standard model.

It would be interesting to see what additional breaking chains discussed in Sec.~VI 
could allow for the confinement scales to approach this ratio without even considering 
differences in the fermion masses between the two sectors.

Since the achievement of gauge coupling constant unification in non-SUSY GUT models is 
somewhat \emph{ad hoc} and, more importantly, suffers from the gauge hierarchy problem, 
we now turn to SUSY models where these problems are absent.

\section{ \bf  Supersymmetric Asymmetric Symmetry Breaking }\label{sec:SUSYASB}
We now develop a supersymmetric
analogue of the model in Sec.~IV, that is an SU(5) theory with scalar fields in the \textbf{10}
and  \textbf{24}.  
In building the supersymmetric potential we will have to introduce 
another chiral supermultiplet
in the $\mathbf{\overline{10}}$ representation,
$Y$, to make it possible to include gauge invariant terms containing $X \sim 10$ in the
superpotential. We must of course also introduce a counterpart field $Y_d$ for the
sake of the discrete symmetry.

This allows for the construction of a potential including all of the fields from 
the non-SUSY case. However, in order to facilitate asymmetric symmetry breaking it
is key that we have both terms that mix the fields under different representations in each
sector and cross terms between the two sectors. This is not possible with the 
set of fields as they are. 
To achieve this we add a singlet scalar superfield $S$ which
transforms into itself under the discrete symmetry.
Doing so allows for the superpotential to generate all of the necessary cross 
terms for asymmetric symmetry breaking through the F-terms of the scalar potential. 
The chiral supermultiplets involved  are then
\begin{eqnarray}
\Phi_v \sim (24,1),  \qquad  X_v \sim (10,1),  \qquad Y_v \sim (\overline{10},1),  \nonumber\\
\Phi_d \sim (1,24),  \qquad  X_d \sim (1,10),  \qquad Y_d \sim (1,\overline{10}) , 
\end{eqnarray}
and
\begin{equation}
 S \sim (1,1).
\end{equation}
The general superpotential
\begin{eqnarray}
W & = & s_1 ( X_v Y_v +  X_d Y_d) + s_2 (\Phi_v \Phi_v + \Phi_d \Phi_d)
+ s_3 (\Phi_v \Phi_v \Phi_v + \Phi_d \Phi_d \Phi_d) + s_4 ( X_d \Phi_d Y_d + X_v \Phi_v Y_v)
    \nonumber \\
  & + & s_5 (\Phi_v \Phi_v S + \Phi_d \Phi_d S) + s_6 (X_v Y_v S  +
 X_d Y_d S)  + s_7 S + s_8 SS + s_9 SSS
\end{eqnarray}
satisfies $SU(5)_v \times SU(5)_d$ gauge invariance and the $\Bbb{Z}_2$ discrete symmetry. The symmetry breaking possibilities with 
this potential are discussed in more detail in Appendix B.

The complete potential has contributions from the F-terms 
of the superpotential,
the D-terms from those fields which are charged under one of the SU(5) symmetries
and soft mass and trilinear terms. Since we have a complete singlet $S$, the non-holomorphic trilinear terms are taken to be absent \cite{Martin:1997ns}. The equation is
\begin{eqnarray}
 V  & = & {W^{i}}^{*} W_i + \frac{1}{2} \sum\limits_a (g\Phi_i T_a \Phi^i)^2  
           - m_X ({X_v}_{ij}{X_v}^{ji}+ {X_d}_{ij}{X_d}^{ji}) \nonumber \\
    & - & m_Y ({Y_v}_{ij} {Y_v}^{ji}+ {Y_d}_{ij} {Y_d}^{ji})   -m_{\Phi} (\Phi_v \Phi_v +\Phi_d \Phi_d ) -m_S S^2 \nonumber \\
    & - &  a_{1}( \Phi_d \Phi_d \Phi_d +   \Phi_v \Phi_v \Phi_v) 
           - a_{2}( X_d Y_d \Phi_d +  X_v Y_v \Phi_v),
\end{eqnarray}
where
\begin{equation}
 W^i=\frac{\partial W}{\partial \phi_i}
\end{equation}
and each $\phi_i$ is one our fields.  
There are nine parameters from the superpotential ($s_1$,...,$s_9$),
six parameters from the soft terms $m_{\Phi}$, $m_X$, $m_Y$, $m_S$, $a_1$, $a_2$ as well as the SU(5) 
coupling constant present in the D-terms.
With this field content we find that the 
scalar potential then has the capacity to display asymmetric symmetry 
breaking by appropriate choice of the parameters. The singlet field $S$ is important here. 
Without it we could not arrive at a scalar potential that has terms 
such as $\Phi_v \Phi_v \Phi_d \Phi_d$, that is, terms which mix the two sectors.
Without these it is not possible to create the necessary dependence between 
sectors for VEV development to be opposing. There are non-minimal choices one could make for the additional fields that
would allow for these terms but for now we choose to simply focus on the simplest case.

Consider a parameter choice with $s_4$ and $s_5$ large compared to
the other superpotential parameters, and with nonzero values of $m_X$, $m_Y$ and $m_{\Phi}$.
F-terms of the style $(\Phi_v^2 \Phi_d^2)$ or $(X_v X_v \Phi_v \Phi_v)$ can then
serve as the cross terms that 
create the asymmetric acquisition of 
VEVs. 
With largely positive quartic terms coming from the D-terms and negative quadratic
terms in the form of the soft masses, these cross terms can 
drive one variety of each multiplet of a given dimensionality to zero in the same manner
as the non-SUSY case. 
It is however the case that many other W terms can spoil this pattern and so many of 
the other superpotential parameters must be kept relatively small, at least an order of magnitude.
The parameter $s_9$ we can allow to be large, as it will serve to bring the value of $S$ to zero.
In one scenario one can generate a nonzero VEV for $\Phi_v$ in the visible sector, 
again breaking 
\begin{equation}
 SU(5)_v \rightarrow SU(3) \times SU(2) \times U(1),
\end{equation}
 and in the dark sector we have $\Phi_d$ developing
a VEV of zero. 
Then the multiplets $X_d$ and $Y_d$ together acquire nonzero VEVs which break 
\begin{equation}
 SU(5)_d \rightarrow SU(3)\times SU(2).
\end{equation}
Being a pair of conjugate representations, they will induce breaking to
the maximal stability 
group of SU(5) according to Michel's conjecture \cite{Michel:1980pc,Slansky:1981yr} which states that this is
the case for a potential containing only a real representation or a pair of conjugate
representation.
This does not strictly apply in this scenario, of course, because we have other fields
involved in the potential. However, we invoke it as numerical analysis shows that 
symmetry breaking of this type occurs within the parameter space that gives asymmetric VEV patterns. 
Appendix B contains further details of this parameter space.
For the \textbf{10} dimensional representation, the maximal stability group, or maximal little 
group, is SU(3) $\times$ SU(2) as it is the only maximal group which observes a singlet within the \textbf{10} of SU(5). 

The supersymmetric case is more constrained in its ability to display asymmetric configurations, 
though with suitable additions in particle content we have found that it is a feature 
that a unified supersymmetric theory can have.
Many of the parameters in the superpotential must be kept quite small so as to not
overpower the terms essential for guaranteeing asymmetric VEV arrays. It would be interesting to 
explore this issue further in developing a complete theory and examining more of the possibilities for
asymmetric SUSY sectors, however that is beyond the scope of this work. 

We now discuss the dependence of
the confinement scale with various parameters in a general supersymmetric theory.

\section{\bf Supersymmetric Confinement} \label{sec:SUSYDarkSU3}
%\noindent{
In the case of supersymmetric theories the running coupling is modified by the additional
particle content. For SU(3) we are however only interested in those particles with colour charge. 
Note that this analysis is not dependent on any 
particular choice of GUT group, relying only on an $SU(3)_v \times SU(3)_d$ structure after GUT breaking.

In the MSSM the one loop beta function for SU(3) is altered by the addition of the gluinos
and sfermions as per
\begin{equation}
\beta_0 =  11- \frac{2}{3}n_f - C_g - \frac{2}{6 }n_s ,%(6 squarks)
\end{equation}
where $n_f$ ($n_s$) is the number of quarks (squarks) and $C_g=2$ is due to the gluinos.
The calculation of the dependence of confinement scale is more model dependent here as one 
must first of all take into account the mass that visible sector 
gluinos and squarks take to consider what value the coupling will take at the GUT or high reference scale $\mu_0$.
This will alter the precise calculation of the value of ${\alpha_3}_d$ at the scale at which 
the visible and dark sector couplings unify.
One can also consider in the dark sector how we might separate the scales of the quarks
and squarks.
If we take the assumption that the SUSY breaking scale is no higher than the mass scale of the dark quarks in the 
dark sector then this provides a rough 
upper bound on the scale at which we place the supersymmetric partners in that sector. This 
assumption is favourable also as it allows for a similar analysis as before in that, 
if the two sectors have SU(3) gauge symmetry with the same number of particles of each kind
all the way down in energy to the mass of the heaviest dark quark, then we can choose this as our
high reference scale and take the value of the coupling at this scale to be the same in both supersymmetric sectors.
Then we can establish a range of possible confinement scales that supersymmetric dark QCD could have. 
We will examine the relationship between the confinement scale and these mass 
scales as we did in the non-SUSY case. In this case we take the squarks and gluinos of the dark sector to be quite light (under a TeV) and in 
such a scenario the dependence is similar to the non-SUSY case but with a larger confinement scale, shown in Fig.\ 3.

\begin{figure}[t!]
\centering
\includegraphics[angle=0,
width=0.7\textwidth]{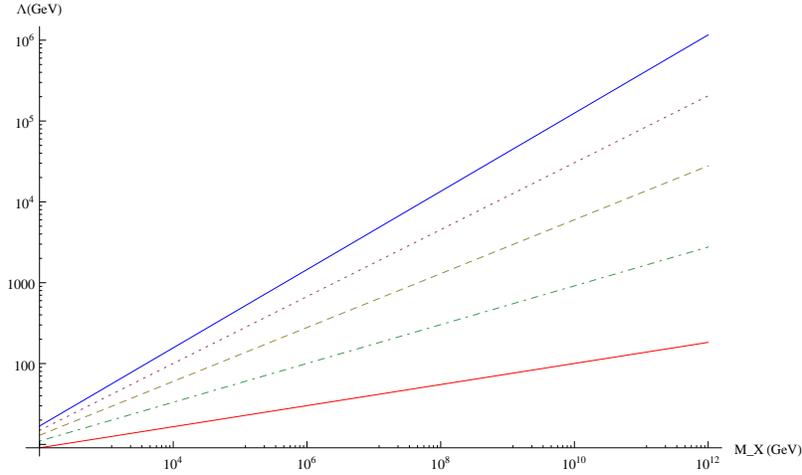}
\caption{Confinement scale dependence on fermion masses, in simple SUSY case, 
almost identical to non-SUSY, but with the confinement scale axis multiplied by $\sim 10$. }
\end{figure}

We now examine the dependence of the dark confinement scale on the dark SUSY breaking scale for a range of different dark-quark masses.
The scale of dark-quark masses is taken to be higher than the SUSY breaking scale in each case. Figures 4-6 show this dependence for different numbers of heavy dark quarks.
The value of the confinement scale is in general higher than the non-SUSY case though we 
do have additional parameters to contend with in the form of the mass scales of the squarks and gluinos.

\begin{figure}[t!]
\centering
\includegraphics[angle=0,
width=0.7\textwidth]{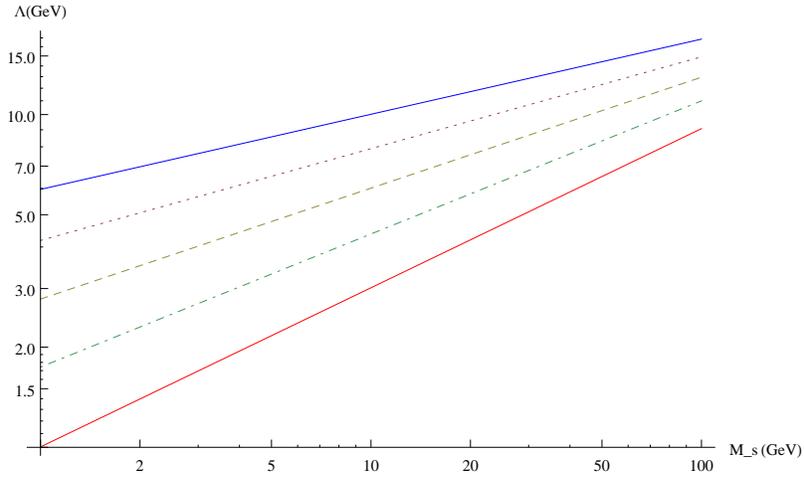}
\caption{Confinement scale dependence on SUSY breaking scale for fixed dark-quark mass scale of 100 GeV. The number of heavy quarks at the dark-quark mass scale ranges from five at the top to one at the bottom.  }
\label{fig:100GeV}
\end{figure}

\begin{figure}[t!]
\centering
\includegraphics[angle=0,
width=0.7\textwidth]{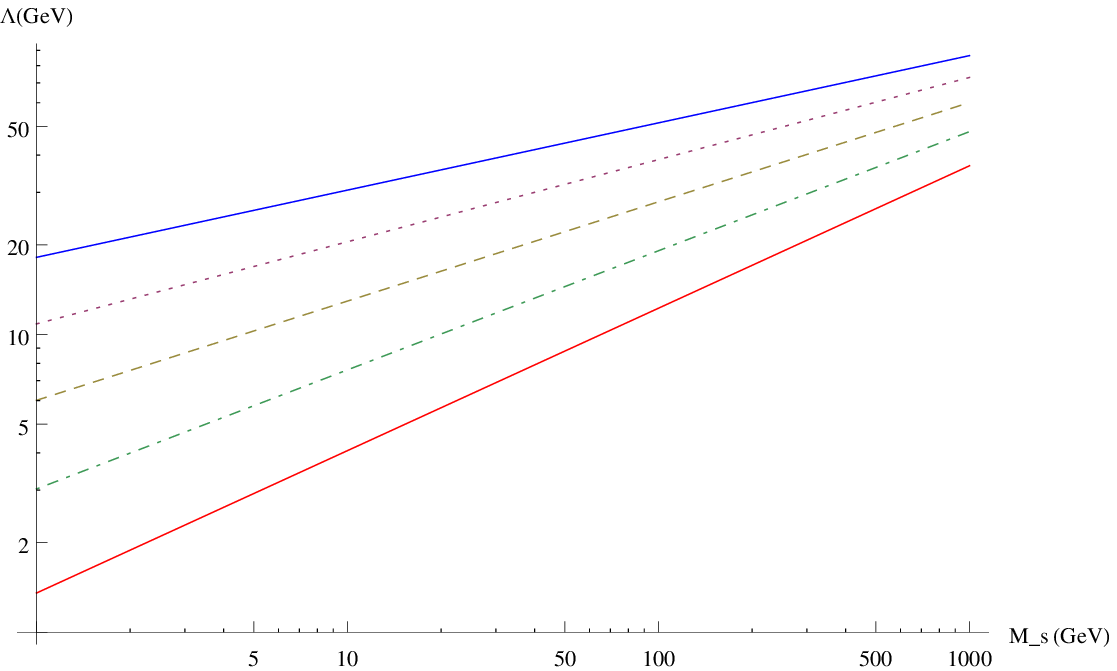}
\caption{As for Fig.\ \ref{fig:100GeV} but with a dark-quark mass scale of 1000 GeV. }
\end{figure}

\begin{figure}[t!]
\centering
\includegraphics[angle=0,
width=0.7\textwidth]{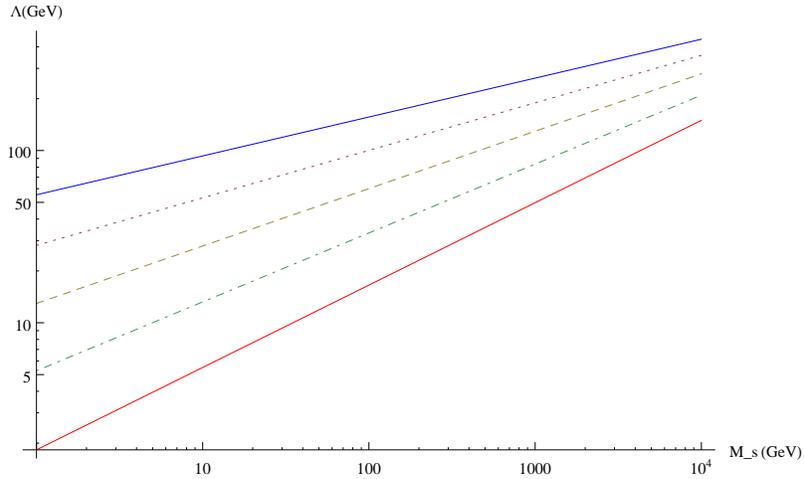}
\caption{As for Fig.\ \ref{fig:100GeV} but with a dark-quark mass scale of $10^4$ GeV.  }
\end{figure}

%----------------------------------------------------------------------------
%CONCLUSION
%----------------------------------------------------------------------------
\section{\bf Conclusions}\label{sec:conc}
Dark matter may be the manifestation of a perhaps quite complicated dark sector that is described by a
gauge theory similar to the standard model.  The cosmological `coincidence' $\Omega_{DM} \simeq 5 \Omega_{VM}$ encourages the thought that DM has a similar origin to VM, and serves as the main
motivation for asymmetric DM.  These models typically succeed in relating the baryon and DM \emph{number} densities, but have nothing very profound to say about the dark matter mass scale.  But 
the latter is as important as the former in considerations of the mass-density `coincidence'.  In most asymmetric DM models, the similar number densities imply that the DM mass should be similar to the proton mass, usually a factor of a few higher.  What could be the origin of such a DM mass scale?

We have explored the idea that grand unification may provide an explanation.  Beginning with a mirror-matter style $G \times G$ gauge group augmented by $\Bbb{Z}_2$ interchange symmetry, we invented a process termed `asymmetric symmetry breaking' which sees the two factors of $G$ break in different ways.  For the asymmetric DM application, we required that both sectors feature unbroken SU(3) subgroups, with different but related confinement scales.  The ordinary QCD confinement scale sets the proton mass, while its dark-sector analogue sets the mass scale of the dark baryon that serves as the DM.  We demonstrated that a significant region of parameter space furnishes a dark confinement scale within an order of magnitude or so above the QCD scale, as favoured by most asymmetric DM models.  Much higher scales are also possible, of course.  We investigated both non-supersymmetric and, more compellingly, supersymmetric GUTs of this type, and in the process explained how dark-quark mass generation can naturally differ from quark mass generation.  Our analysis serves as a starting point for building fully-realistic asymmetric DM models from a grand unification base.

The possibilities inherent in asymmetric symmetry breaking are rather large when one considers $G =$ SO(10) and other higher-rank options.  This seems to offer fruitful avenues for future investigations, and may ultimately serve to provide a truly unified understanding of the microphysics and macrophysics of ordinary and dark matter.

\acknowledgments{We thank Michael A. Schmidt for comments on an earlier draft. SJL thanks B. Callen and A. Sharma for helpful discussions. This work was supported in part by the Australian Research Council.}

\newpage 
\appendix
\section{\bf Scalar potential for non-supersymmetric SU(5)$\times$SU(5) model}\label{sec:AppA}
In Sec.~IV we outlined the construction of an SU(5)$\times$ SU(5) potential with asymmetric minima. Here we discuss its features in more detail and explore some of the possibilities in regard to breaking to various subgroups.
The full SU(5) $\times$ SU(5) potential can be written as
\begin{eqnarray}
V & = & \lambda_{a1}({\phi_v}^i_j{\phi_v}^j_i + {\phi_d}^i_j{\phi_d}^j_i - \mu_a^2)^2 +  
                    \kappa_a ({\phi_v}^i_j{\phi_v}^j_i {\phi_d}^h_k{\phi_d}^k_h) + 
                    \lambda_{a2}( {\phi_v}^i_j {\phi_v}^j_k {\phi_v}^k_h {\phi_v}^h_i           +            {\phi_d}^i_j{\phi_d}^j_k {\phi_d}^k_h {\phi_d}^h_i ) \nonumber \\
          & + &          \lambda_{t2}(  {\chi_v}_{ij}{\chi_v}^{ji}+ {\chi_d}_{ij}{\chi_d}^{ji}- \mu_t^2)^2 +
                    \kappa_t ({\chi_v}_{ij}{\chi_v}^{ji} {\chi_d}_{ij}{\chi_d}^{ji} )\nonumber \\
           & + &         \lambda_{t2} ( {\chi_v}_{ij}{\chi_v}^{ij}{\chi_v}_{ij}{\chi_v}^{ij}          +              {\chi_d}_{ij}{\chi_d}^{ij}{\chi_d}_{ij}{\chi_d}^{ij}) +\                   
                 C_0 (  {\chi_v}_{ij}{\chi_v}^{ij}{\phi_v}^i_j {\phi_v}^j_i                          +                {\chi_d}_{ij}{\chi_d}^{ji}{\phi_d}^i_j {\phi_d}^j_i) \nonumber \\
         & + &        C_1 (  {\chi_v}_{ij}{\chi_v}^{jk}{\phi_v}^l_k {\phi_v}^j_l                          +                {\chi_d}_{ij}{\chi_d}^{jk}{\phi_d}^l_k {\phi_d}^j_l) +
                 C_2 (  {\chi_v}_{lq}{\chi_v}^{ij}{\phi_v}^l_j {\phi_v}^q_i                          +               {\chi_d}_{lq}{\chi_d}^{ij}{\phi_d}^l_j {\phi_d}^q_i  ) \nonumber \\
      & + &        C_3 (  {\chi_v}_{su}{\chi_v}^{pq}{\phi_v}^n_m {\phi_v}^i_j   \epsilon^{smujt}  \epsilon_{pnqit}      +
                         {\chi_d}_{su}{\chi_d}^{pq}{\phi_d}^n_m {\phi_d}^i_j   \epsilon^{smujt}  \epsilon_{pnqit}   ) \nonumber\\
          & + &     C_4 (    {\phi_v}^i_j{\phi_v}^j_i {\chi_d}_{ij}{\chi_d}^{ji}
          +                {\phi_d}^i_j{\phi_d}^j_i {\chi_v}_{ij}{\chi_v}^{ji}  ).
\end{eqnarray}
The parameters are $\lambda_{t1},\mu_{t}, \lambda_{t2}$ as well as $\lambda_{a1},\mu_{a}, \lambda_{a2}$, $\kappa_{a}$ and $\kappa_{t}$.
In addition to these there are five cross terms arising from nontrivial 
contractions between our representations, with parameters $(C_0,C_1,C_2,C_3,C_4)$.
In general the asymmetry required can be attained by making these additional
cross term parameters smaller than $C_0$ and the other parameters of the model.
In minimising this potential we can reduce the total number of parameters by placing all of our fields in a simplified VEV form. The adjoint can be represented by the traceless matrix
\begin{equation}
\braket{\phi_v}= v_v
\begin{pmatrix}
  \alpha_1 & 0 & 0 & 0 & 0 \\
   0 & \alpha_2 & 0 & 0 & 0 \\
 0 & 0 & \alpha_3 & 0 & 0 \\
  0 & 0 & 0 & \alpha_4 & 0 \\
   0 & 0 & 0 & 0 & \alpha_5 \\
 \end{pmatrix},
\end{equation}
with $\alpha_1 +\alpha_2 +\alpha_3 +\alpha_4 +\alpha_5 =0 $. For the \textbf{10} we have
\begin{equation}
\braket{\chi_d} = v_d
 \begin{pmatrix}
  0 & \rho_1 & 0 & 0 & 0 \\
  -\rho_1 & 0 & 0 & 0 & 0 \\
  0 & 0 & 0 & \rho_2 & 0 \\
  0 & 0 & -\rho_2 & 0 & 0 \\
  0 & 0 & 0 & 0 & 0 \\
 \end{pmatrix},
\end{equation}
with $\rho_{1,2}$ complex. The  \textbf{24}  and \textbf{10} are both reduced to just four total different degrees of freedom each in this form. 
Working numerically we can however quickly compare the results of using just these 16 degrees of freedom or the full 68; they were found to agree in all cases.
The parameter space is directly comparable to that of the simple model of Sec.~III. The positive definite terms act exactly like collections of additional 
fields that one could add to that previous model with the same-sector and cross-sector couplings needed to generate asymmetric VEVs that differentiate entire sets 
of fields within these multiplets. That is, if $\kappa_a$ is large enough then if all (${\phi_v}^i_j$) fields gain a nonzero VEV, all of the fields  (${\phi_d}^i_j$) are encouraged to become zero.
Together with ($C_1,C_2,C_3,C_4$) there is a greater variability for the signs of quartic terms of the potential. Scaling any of these additional quartics too high may alter the VEV pattern from the desired asymmetric pattern.
A larger value of $C_0$ will however ensure the breaking is the extension of that in Sec.~III.
To be concrete, we display an example of some parameters set along these guidelines and the VEVs that are produced.  The parameters
\begin{eqnarray}
& \lambda_{a1} \simeq 0.4, \     
\kappa_{a} \simeq 0.4, \     
\kappa_{t}  \simeq 0.4, \
\lambda_{t1} \simeq 0.8, & \nonumber \\    
& \mu_{t}  \simeq 0.2, \    
\mu_{a}  \simeq 0.1, \     
\lambda_{a2}  \simeq 0.1, \      
\lambda_{t2}  \simeq -0.1, & \nonumber \\    
& C_0  \simeq 0.5, \    
C_1  \simeq -0.1,  \ 
C_2  \simeq -0.1, \
C_3  \simeq -0.1,  \    
C_4  \simeq -0.1 &
\end{eqnarray}
give rise to the VEVs
\begin{eqnarray}
\braket{\phi_v} & \simeq & 0.24
\begin{pmatrix}
  1 & 0 & 0 & 0 & 0 \\
   0 & 1 & 0 & 0 & 0 \\
 0 & 0 & 1 & 0 & 0 \\
  0 & 0 & 0 & -3/2 & 0 \\
   0 & 0 & 0 & 0 & -3/2 \\
 \end{pmatrix},
\nonumber \\
\braket{\chi_v} & \simeq & 0, \nonumber \\
\braket{\phi_d} & \simeq & 0, \nonumber \\
\braket{\chi_d} & \simeq &0.1
 \begin{pmatrix}
  0 & 1+i & 0 & 0 & 0 \\
  -1-i & 0 & 0 & 0 & 0 \\
  0 & 0 & 0 & 0 & 0 \\
  0 & 0 & 0 & 0 & 0 \\
  0 & 0 & 0 & 0 & 0 \\
 \end{pmatrix}.
\end{eqnarray}

\section{ \bf Scalar potential for supersymmetric SU(5)$\times$SU(5) model}\label{sec:AppB}
In this section we will discuss further the results of the supersymmetric version of asymmetric symmetry breaking.
The analysis here only serves to demonstrate that such asymmetric patterns are possible within the constraints inherent in supersymmetric theories.

Positive definite couplings between fields of different sectors are required to create the anti-correlation between sectors. This is what necessitates a field which transforms into itself under
the discrete symmetry. An alternative to this could be to arm the theory with a pair of complete singlets under the discrete symmetry, i.e.\ $S_v$, $S_d$. 
Without such additions we are unable to create gauge invariant terms in the superpotential which can allow for cross-sector couplings to appear in the F-terms. 
The other addition we made of the multiplet $Y$ was based on our choice of complex representations.\footnote{This may of course not be necessary, if one was working with two different real representations to facilitate different symmetry breaking in each sector. 
In that case the procedure would be more straightforward.} 
We wish, however, to demonstrate that the theory which we used previously can be adopted into a supersymmetric form with the same gauge group breaking chains.
The terms that we wish to highlight that are derived from the superpotential are the contractions of the form 
\begin{equation}
s_4^2 (\Phi_v \Phi_v X_v X_v + \Phi_v \Phi_v Y_v Y_v + \Phi_d \Phi_d X_d X_d + \Phi_d \Phi_d Y_d Y_d). 
\end{equation}
It is clear that the parameter $s_4$ being larger can  help lead to asymmetric VEVs. 
The other important parameter is $s_5$ which affects the term 
\begin{equation}
s_5^2 (\Phi_d \Phi_d \Phi_v \Phi_v).
\end{equation} 
With just these terms and the additional soft masses one can generate an asymmetric VEV pattern.
For the parameter example
\begin{eqnarray}
& s_4 = s_5 \simeq 0.02, \qquad g_5 \simeq 0.037, \qquad s_9 \simeq 0.001, & \nonumber \\
& m_X = m_Y \simeq 0.001, \qquad m_{\Phi} \simeq 0.1, \qquad m_S = 0, &
\end{eqnarray}
and all trilinear terms and other parameters set at or close to zero, we obtain nonzero VEVs for the adjoint in one sector and for the fields $X_v$ and $Y_d$ in the other sector which serve to break $SU(5)_v$ 
to the standard model gauge group and $SU(5)_d$ to the dark sector gauge group with VEVs
\begin{eqnarray}
\braket{\Phi_v} & \simeq & 2.1
\begin{pmatrix}
  1 & 0 & 0 & 0 & 0 \\
   0 & 1 & 0 & 0 & 0 \\
 0 & 0 & 1 & 0 & 0 \\
  0 & 0 & 0 & -3/2 & 0 \\
   0 & 0 & 0 & 0 & -3/2 \\
 \end{pmatrix},
\nonumber \\
\braket{X_v} & \simeq & 0, \nonumber \\
\braket{Y_v} & \simeq & 0, \nonumber \\
\braket{\Phi_d} & \simeq & 0, \nonumber \\
\braket{S} & \simeq & 0, \nonumber \\
\braket{X_d} & \simeq &
\begin{pmatrix}
  0 & 1.2+2.9i & 0 & 0 & 0 \\
  -1.2-2.9i & 0 & 0 & 0 & 0 \\
  0 & 0 & 0 & -1.53-2.1i & 0 \\
  0 & 0 & 1.5+2.1i  & 0 & 0 \\
  0 & 0 & 0 & 0 & 0 \\
\end{pmatrix},
\nonumber \\
\braket{Y_d} & \simeq &
\begin{pmatrix}
  0 & 1.2-2.9i & 0 & 0 & 0 \\
  -1.2+2.9i & 0 & 0 & 0 & 0 \\
  0 & 0 & 0 & -1.53+2.1i & 0 \\
  0 & 0 & 1.5i-2.1i  & 0 & 0 \\
  0 & 0 & 0 & 0 & 0 \\
\end{pmatrix}.
\end{eqnarray}

This demonstrates the capacity for supersymmetric models to display the same asymmetric symmetry breaking as non-SUSY models.
There are other terms which can contribute to the asymmetric pattern, i.e. contractions of the style $(X_d X_d X_v X_v)$, but scaling these up to be larger also scales upwards terms that we would need to contend with 
to maintain the asymmetry.

\bibliography{grandunifieddarkmatterbib.bib} % Bibliography -

\end{document}